\documentclass[useAMS,usenatbib]{mn2e}
\usepackage{epsfig,graphicx}

\title[Active Region Evolution]{Plasma Diagnostics of Active Region Evolution and Implications for Coronal Heating}
\author[R. Milligan et al.]{R. O. Milligan,$^{1,3}$ \thanks{E-mail: R.Milligan@qub.ac.uk} P. T. Gallagher,$^{2,3,4}$ M. Mathioudakis,$^1$ F. P. Keenan,$^1$
\newauthor 
and D. S. Bloomfield,$^1$\\
$^1$Department of Physics and Astronomy, 
      Queen's University Belfast, University Road,
      Belfast, BT7 1NN, Northern Ireland\\
$^2$Department of Experimental Physics, University College Dublin,
      Belfield, Dublin 4, Ireland\\
$^3$Laboratory for Astronomy and Solar Physics, 
      NASA Goddard Space Flight Centre, Greenbelt, MD 20771, U.S.A.\\
$^4$L-3 Communications GSI\\}
\begin{document}

\date{Released 2005 Xxxxx XX}

\pagerange{\pageref{firstpage}--\pageref{lastpage}} \pubyear{2005}

\label{firstpage}

\maketitle

\begin{abstract}
A detailed study is presented of the decaying solar active region NOAA 10103 
observed with the Coronal Diagnostic Spectrometer (CDS), the Michelson Doppler
Imager (MDI) and the Extreme-ultraviolet Imaging Telescope (EIT) onboard the {\it Solar and Heliospheric Observatory (SOHO)\/}. Electron density maps formed using
\textsc{Si x}(356.03~\AA/347.41~\AA) show that the density varies from
$\sim$10$^{10}$~cm$^{-3}$ in the active region core, to
$\sim$7$\times$10$^{8}$~cm$^{-3}$ at the region boundaries. Over the five days of
observations, the average electron density fell by $\sim$30\ per cent.
Temperature maps formed using \textsc{Fe xvi}(335.41~\AA)/\textsc{Fe xiv}(334.18~\AA) show electron temperatures of $\sim$2.34$\times$10$^{6}$~K in the active region core,
and $\sim$2.10$\times$10$^{6}$~K at the region boundaries. Similarly to the 
electron density, there was a small decrease in the average electron  temperature
over the five day period. The radiative, conductive, and mass flow losses were calculated
and used to determine the resultant
heating rate ($P_{H}$). Radiative losses were found to dominate the
active region cooling process. As the region decayed, the heating rate decreased
by almost a factor of five between the first and last day of observations. The heating rate was then compared to the total unsigned magnetic
flux ($\Phi_{tot} = \int dA |B_z|$), yielding a power-law of the form  $P_{H}\sim\Phi_{tot}^{0.81\pm0.32}$. This result suggests that waves rather than nanoflares may be the dominant heating mechanism in this active region.
\end{abstract}


\section{INTRODUCTION}
\label{intro}

Since the discovery of highly ionized species of iron in the solar corona in
the 1930's, physicists have been puzzled by the high temperatures observed in
the outer solar atmosphere \citep{edlen37}. It is widely accepted that the 
magnetic field plays a fundamental role in the heating process, but precise
measurements of the coronal magnetic field are currently impossible. Indirect 
methods are therefore adopted which rely on measurable quantities such as the 
electron temperature, density and photospheric magnetic flux. These measurements
can then be compared to theoretical predictions. 

Models for coronal heating typically belong to one of two broad categories. In wave (AC) 
heating, the large-scale magnetic
field essentially acts as a conduit for small-scale, high-frequency Alfv\`{e}n
waves propagating into the corona. For constant Alfv\`{e}n wave amplitude
$\langle v^{2} \rangle$, the total power dissipated in an active region is,
\begin{equation}
\label{ac_heating}
P_H = \sqrt{\frac{\rho}{4\pi}} \langle v^2 \rangle \Phi_{tot}~~{\rm~ergs~s^{-1}},
\end{equation}
where $\rho$ is the mass density, and $\Phi_{tot}$ is the total unsigned magnetic flux, 
\begin{equation}
\label{phi_tot}
\Phi_{tot} = \int dA |B_z|~~{\rm~Mx},
\end{equation}
where $B_z$ is the longitudinal component of the magnetic field.

\begin{figure*}
\includegraphics[height=16.5cm,angle=90]{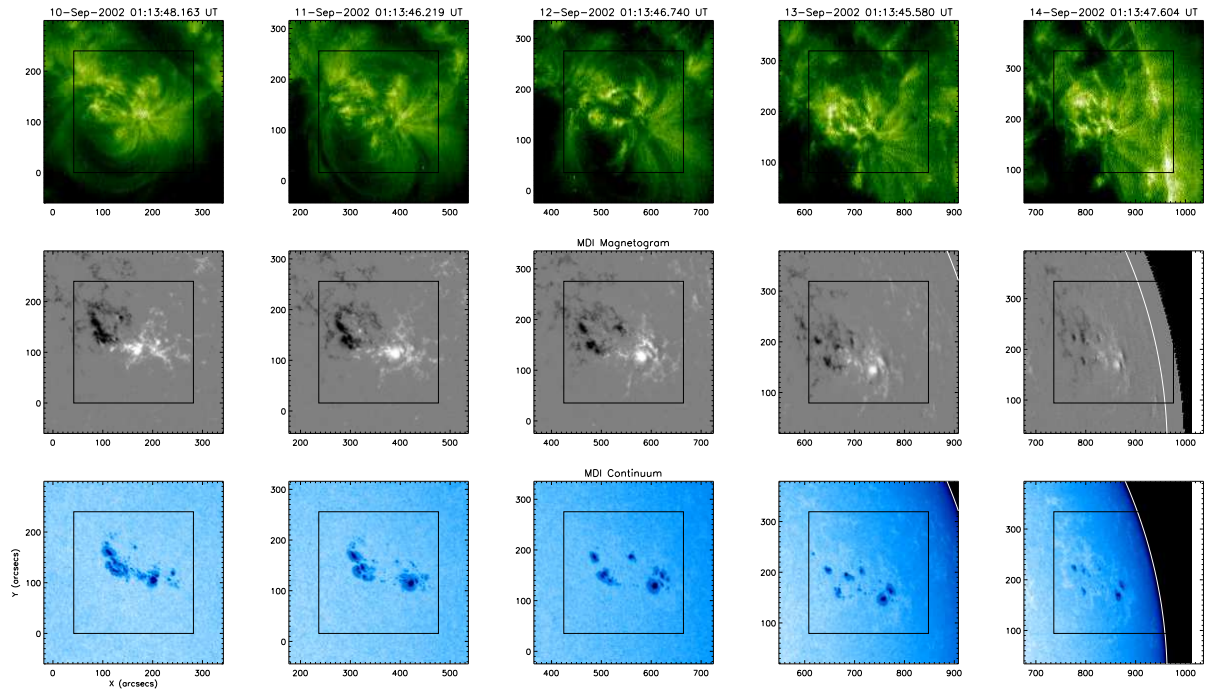}
\caption{The evolution of NOAA 10103 during September 10--14. {\em Top row:} EIT (195~\AA) images. {\em Middle row:} MDI magnetograms. {\em Bottom row:} MDI continuum images. The inner boxes in each image denote the field-of-view of the CDS raster.}
\label{eit_mdi}
\end{figure*}

In stress (DC) heating, the coronal magnetic field stores energy in the form of 
electric currents until it can be dissipated, e.g., by nanoflares
\citep{park88}. The total power can be estimated by,
\begin{equation}
\label{dc_heating}
P_H \sim |v| \Phi_{tot}^{2}~~{\rm~ergs~s^{-1}},
\end{equation}
and the constant of proportionality describes the efficiency of magnetic
dissipation, which might involve the random footpoint velocity, $v$
\citep{park83}, or simply the geometry \citep{brow86,fish98}.

Several authors have linked the photospheric magnetic flux to EUV and X-ray line
intensity. \citet{gurm74} found that the line intensity of \textsc{Mg x} (624.94~\AA) was proportional to the magnetic
flux density. \citet{schr87} related the integrated intensities of
chromospheric, transition region, and coronal lines to the total magnetic flux
by a power law, the index of which was dependent on the scale height. This
result was later confirmed by \citet{fish98}, who showed that X-ray luminosity
is highly correlated with the total unsigned magnetic flux. Van Driel-Gesztelyi
et al.~(2003) also showed a power-law relationship between the mean X-ray flux,
temperature, and emission measure, and the mean magnetic field by studying the
long-term evolution of an active region over several rotations, at times when 
there were no significant brightenings. 

In this paper, we study the evolution of a decaying active region using the
diagnostic capabilities of the Coronal Diagnostic Spectrometer (CDS; Harrison et
al.~1995) and Michelson Doppler Imager (MDI; Scherrer et al.~1995) onboard {\it SOHO}. 
Using the temperatures, densities, and dimensions of the active region, the heating 
rate is calculated and compared to the total unsigned magnetic flux. 
These results can be put in the context of theoretical models.
Section~\ref{obs} gives a brief overview of the active region, a summary of the 
instruments involved, and a description of the 
data analysis techniques. Our results are given in Section~\ref{results},  
and discussion and conclusions in Section~\ref{conc}.

\section{OBSERVATIONS AND DATA ANALYSIS}
\label{obs}

NOAA 10103\footnotemark[1]\footnotetext[1]{See
http://www.solarmonitor.org/20020910/0103.html} was observed by CDS, EIT,
and  MDI for five consecutive days during 2002 September 10--14. Fig.~\ref{eit_mdi} 
shows the general evolution of the region over that period. 

\begin{figure*}
\includegraphics[width=16.5cm]{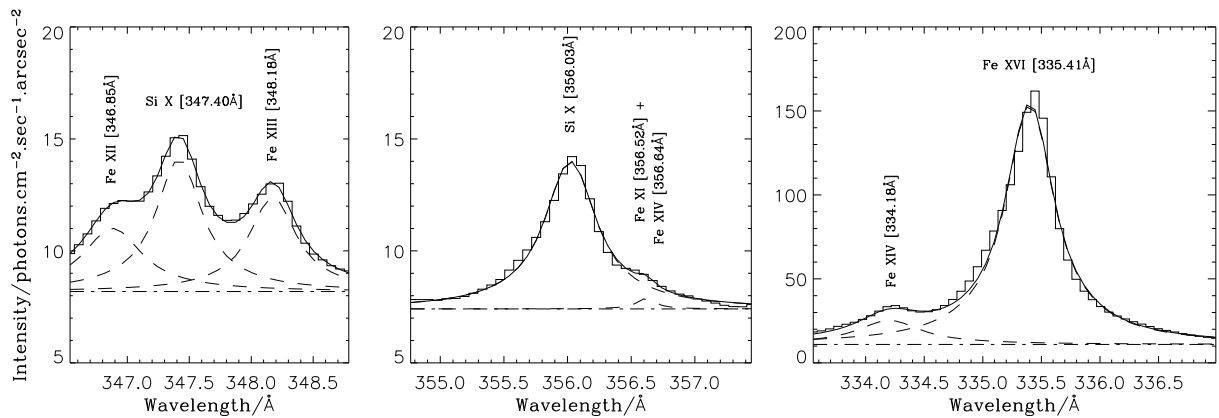}
\caption{Portions of the CDS spectra from NOAA 10103 on September 10, together with backgrounds (horizontal dot-dash lines) and broadened Gaussian fits.}
\label{line_fits}
\end{figure*}

\subsection{The Coronal Diagnostic Spectrometer}
\label{CDS}

EUV spectra were obtained with the CDS instrument, which is a dual
spectrometer that can be used to obtain images with a spatial resolution of
$\sim$8 arcsec. The Normal Incidence Spectrometer (NIS), used in this study,
is a stigmatic spectrometer which forms images by moving the solar image across
the slit using a scan mirror. The spectral ranges of NIS (308--381~\AA~and
513--633~\AA) include emission lines formed over a wide range of temperatures,
from $\sim$10$^{4}$~K at the upper chromosphere, through the transition region,
to $\sim$10$^{6}$~K at the corona. The details of the \textsc{az\_ddep1} observing sequence
used in this study can be found in Table~\ref{CDS_table}. 

\begin{table}
\small
\caption{Details of the CDS \textsc{az\_ddep1} observing sequence}
\label{CDS_table}
\begin{tabular}{lc} 
\hline
\multicolumn{1}{c}{\small{Parameter}}	&\small{Value}\\
\hline
Date				 &    2002 September 10--14\\
Region name			 &    NOAA 10103\\
Instrument			 &    CDS/NIS1\\
Wavelength range (\AA) 	 	 &    332--368\\
Slit size (arcsec$^{2}$)	 &    4.064$\times$240\\
Area imaged (arcsec$^{2}$)	 &    243.84$\times$240\\
Exposure time (s)		 &    50\\
Number of slit positions	 &    60\\
\hline
\normalsize
\end{tabular}
\end{table}

The raw CDS data were cleaned to remove cosmic rays, and calibrated to remove the
CCD readout bias and convert the data into physical units of
photons~cm$^{-2}$~s$^{-1}$~arcsec$^{-2}$.  Due to the
broadened nature of post-recovery CDS spectra, the emission lines were fitted with
modified Gaussian profiles as described by \citet{thom99}.
The Gaussian term was defined as,
\begin{equation}
G(\lambda)=exp\left[-\frac{1}{2}\left(\frac{\lambda-\lambda_0}{\sigma}\right)^2\right],
\end{equation}
and the wings by,
\begin{equation}
W(\lambda)=\frac{1}{\left(\frac{\lambda-\lambda_0}{\sigma'}\right)^2+1},
\end{equation}
where $\lambda$ is the wavelength, $\lambda_0$ is the central wavelength
of the line, $\sigma$ is the Gaussian width, and $\sigma'$ is the FWHM.

The combined function describing the line profile can be expressed as,
\begin{equation}
B(\lambda)=I[(1-a)G(\lambda)+a W(\lambda)],
\end{equation}
where $I$ is the amplitude of the line profile and $a$ can be the amplitude of the red or
the blue wing. The line flux is then given by,
\begin{eqnarray}
\label{flux_eqn} 
\int B(\lambda) d\lambda & = & I \sigma \left[\left(1-\frac{a_{red}}{2}
\left(1+\frac{a_{blue}}{a_{red}}\right) \right) \sqrt{2\pi}\right.\nonumber  \\
                    & + & \left.a_{red} \left(1+\frac{a_{blue}}{a_{red}}\right) \pi
		    \sqrt{2\ln(2)} \right].
\end{eqnarray}

These broadened Gaussian profiles were then fitted to emission lines in three 
wavelength intervals using the \textsc{xcfit} routine in the CDS branch of the 
SolarSoftWare tree (SSW; Freeland \& Handy~1998; see Fig.~\ref{line_fits}). Two of the intervals were centreed on each of the density sensitive \textsc{Si x} (347.41~\AA) and \textsc{Si x} (356.03~\AA) lines, while the third contained the temperature sensitive \textsc{Fe xiv} (334.18~\AA) and \textsc{Fe xvi} (335.41~\AA) pair. Due to the relatively low intensity of \textsc{Fe xiv} (334.18~\AA) compared to that of the adjacent \textsc{Fe xvi} (335.41~\AA) transition, an upper constraint on the width of $\sigma$ = 0.4~\AA~(FWHM = 0.17~\AA) was placed on the \textsc{Fe xiv} line. The primary lines, their formation temperatures, transitions, and rest wavelengths are given in Table~\ref{line_data}.

\begin{table}
\small
\caption{Ions, Formation Temperatures, Transitions and Wavelengths of Emission Lines Identified in This Work}
\label{line_data}
\begin{tabular}{lclc} 
\hline
\multicolumn{1}{c}{Ion}	&Log T$_{e}$	&Transition					&$\lambda$/\AA\\
\hline
\textsc{Si x}	&6.1	&2s$^{2}$2p$^{2}$~$^{3}$P$_{1/2}$--2s2p$^{2}$~$^{2}$D$_{3/2}$	&347.41\\
\textsc{Si x}	&6.1	&2s$^{2}$2p~$^{3}$P$_{3/2}$--2s2p$^{2}$~$^{2}$D$_{3/2,5/2}$  	&356.03\\
\textsc{Fe xiv}	&6.3	&3s$^{2}$3p~$^{2}$P$_{1/2}$--3s3p$^{2}$~$^{2}$D$_{3/2}$		&334.18\\
\textsc{Fe xvi}	&6.4	&3s~$^{2}$S$_{1/2}$--3p~$^{2}$P$_{3/2}$				&335.41\\
\textsc{Fe xvii}&6.7	&2p$^{5}$~3s~$^{3}$P$_{1}$--2p$^{5}$~3p~$^{1}$D$_{2}$		&347.85\\
\hline
\normalsize
\end{tabular}
\end{table}

\subsection{The Michelson Doppler Interferometer}
\label{MDI}

Magnetic field measurements were taken by the MDI instrument, which
images the Sun on a 1024~$\times$~1024 pixel CCD camera through a series of increasingly
narrow filters. The final elements, a pair of tunable Michelson interferometers,
enable MDI to record filtergrams with a FWHM bandwidth of 94~m\AA. Several 
times each day, polarizers are inserted
to measure the line-of-sight magnetic field. In this paper, 5-minute-averaged
magnetograms of the full disk were used, with a 96 minute cadence and a pixel
size of 2 arcsec.

\citet{berg03} analyzed Advanced Stokes Polarimeter (ASP) and MDI
magnetograms, and found that MDI underestimates the magnetic flux 
densities by a
factor of 1.45 for values below 1200~G. For flux densities
higher than 1200~G, this underestimation becomes nonlinear, with 
the MDI
fluxes saturating at $\sim$1300~G. Values below 1200~G were 
therefore corrected by multiplying by 1.45, and values above 
1200~G were approximately corrected by multiplying by a factor of 1.9 \citep{gree03}.

Before calculating the total unsigned magnetic flux, two final corrections were applied to the data. The
first results from the fact that the measured line-of-sight flux
deviates more and more from a radial measurement as one approaches the
limb. For simplicity, we assume that magnetic fields in the
photosphere are predominantly radial. Then, the radial field strength
becomes equal to the line-of-sight field strength times
$1/\cos{\theta}$, where $\theta$ is the heliocentric distance of
the region from Sun centre. 

Active region areas, $A$, were calculated by counting all pixels above 500~G and
multiplying by the appropriate factor to obtain the active region area in cm$^2$.
This threshold value for the magnetic field was found to adequately separate active
region structures from neighbouring areas of quiet-Sun and plage. As the region
approached the limb, the effects of foreshortening became significant. Measured
areas were therefore corrected by dividing by $\cos{\theta}$. The resulting total
unsigned flux was then calculated using Equation~(\ref{phi_tot}). 

Modern high resolution space based instruments, such as the {\it Transition Region and Coronal Explorer (TRACE)}, have been used by \cite{asch01b} to show that coronal loops observed in different bandpasses are not necessarily cospatial. Due to the coarse resolution of CDS, we make the assumption that active regions occupy a hemispherical volume, $V = 2/3~\pi^{-1/2}A^{3/2}$, with a mean loop length of $L = \sqrt{ \pi A}$.

\begin{figure}
\includegraphics[width=8.0cm]{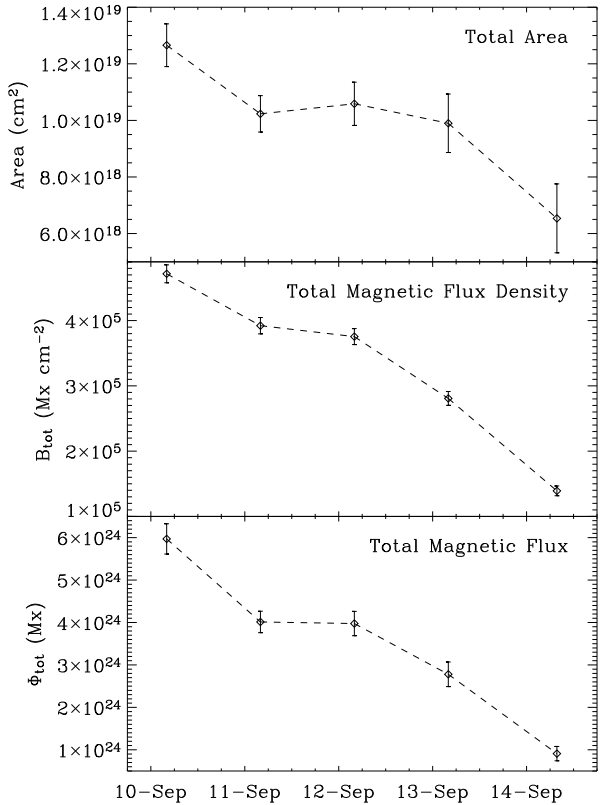}
\caption{The cosine-corrected active region area, total 
magnetic flux density, and total magnetic flux for September 10--14.}
\label{mag_plot}
\end{figure}

\section{RESULTS}
\label{results}

\subsection{Morphology and Magnetic Field}

The top row of Fig.~\ref{eit_mdi} shows a series of 360 $\times$360 arcsec 
EIT images obtained in the 195~\AA~bandpass. The first three images show
a number of loops to the south and north of  the region, which are not visible from
2002 September 13 onwards. On September 10, the MDI magnetogram shows a simple
bipolar $\beta$ region, which is then classified as a $\beta\gamma$ on the
following day. The region subsequently decreased in both size and complexity as it
approached the west limb. The overall decay of the active region is clearest in
the MDI continuum images in the bottom row of Fig.~\ref{eit_mdi}.

Table~\ref{res_tab} and Fig.~\ref{mag_plot} show the decay of the region in terms of the cosine-corrected 
area and magnetic flux for September 10--14. The region was observed to have an
initial area of 1.26$\times$10$^{19}$~cm$^2$, which fell to
6.54$\times$10$^{18}$~cm$^2$ by September 14. The total unsigned magnetic
flux density also shows a similar trend, falling from close to
4.80$\times$10$^{5}$~Mx~cm$^2$ to 1.40$\times$10$^{5}$~Mx~cm$^2$. The product of the region area times the
total magnetic flux density is then given in the bottom panel of Fig.~\ref{mag_plot}. As the
region decays, the total unsigned magnetic flux falls off by a factor of 5--6 over 
the five days from September 10 to 14.

\subsection{Electron Densities}
\label{den_map}

\begin{figure}
\includegraphics[height=8.0cm, angle=90]{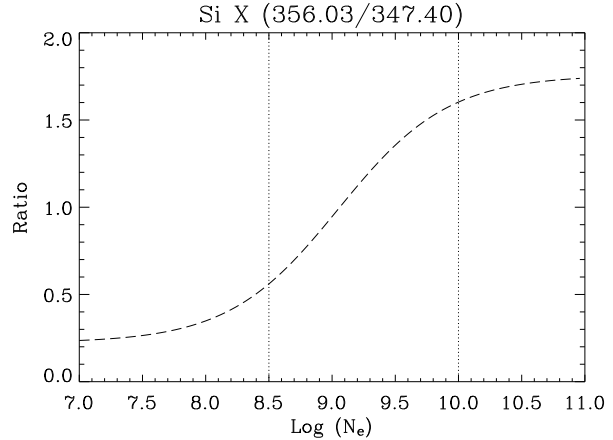}
\caption{Theoretical line ratio as a function of density from \textsc{chianti} for 
\textsc{Si x} (356.03~\AA/347.41~\AA). The vertical dotted lines denote 
the typical range of densities expected to be found in an active region.}
\label{chianti}
\end{figure}

Electron density maps were generated using the \textsc{Si x} (356.03~\AA/347.41~\AA) ratio in conjunction with theoretical data from the \textsc{chianti} v4.2
atomic database \citep{dere97}. Fig.~\ref{chianti} shows a plot of the density
sensitive \textsc{Si x} (356.03~\AA/347.40~\AA) ratio together with the expected
range of densities in an active region of this size and class. 

\begin{figure*}
\includegraphics[width=16.5cm]{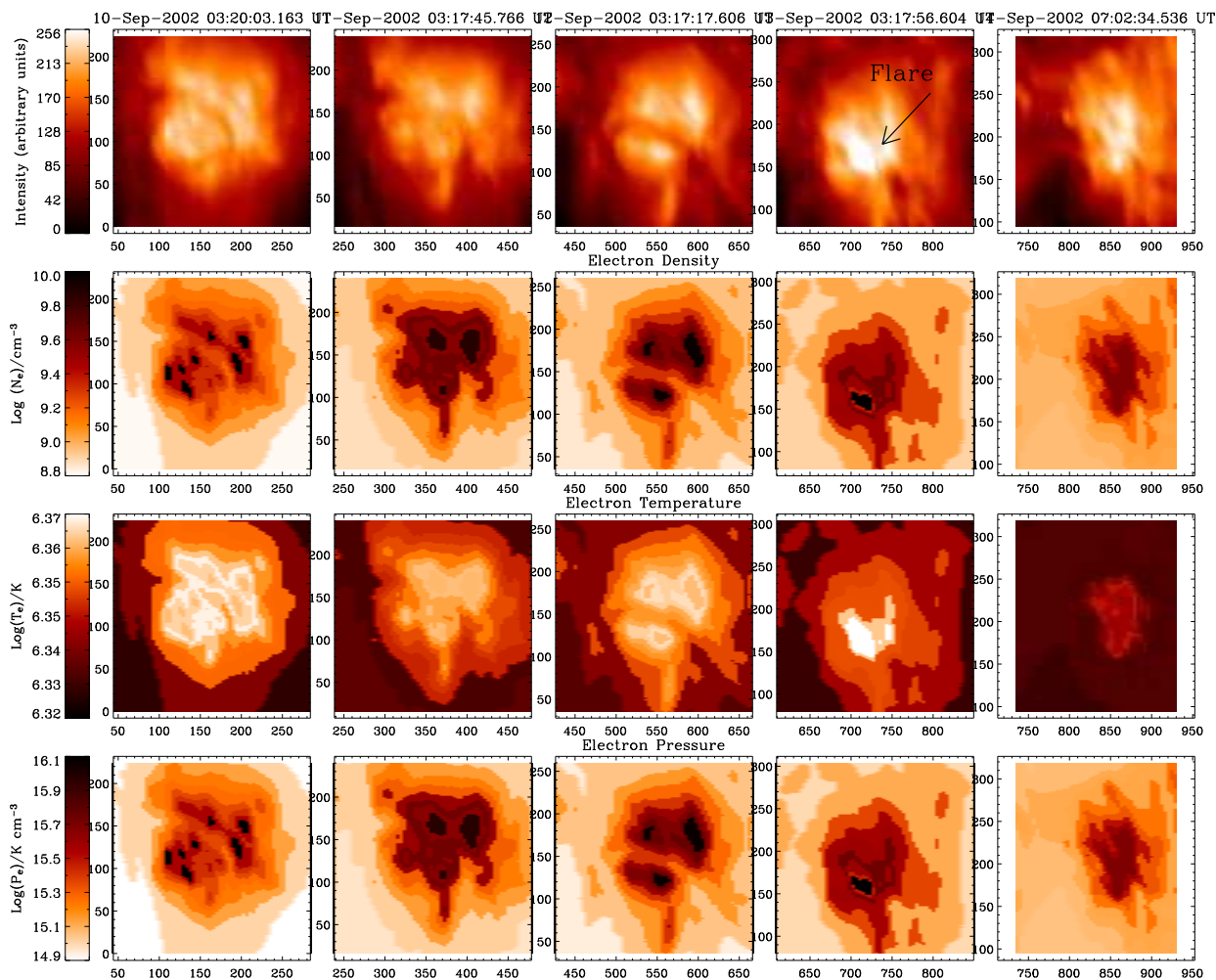}
\caption{{\em Top row:} CDS images of NOAA 10103 over the five days of observations. 
{\em Second row:} Electron density maps derived from the \textsc{Si x} (356.03~\AA/347.41~\AA) ratio. 
{\em Third row:} Electron temperature maps found from the \textsc{Fe xvi} (335.41~\AA)/\textsc{Fe xiv} (334.19~\AA) ratio. 
{\em Bottom row:} Electron pressure maps from the product of the density maps with the formation temperature of the \textsc{Si x} lines. 
The panels for September 13 show the compact brightening.}
\label{maps_fig}
\end{figure*}

\begin{figure}
\includegraphics[width=8.0cm]{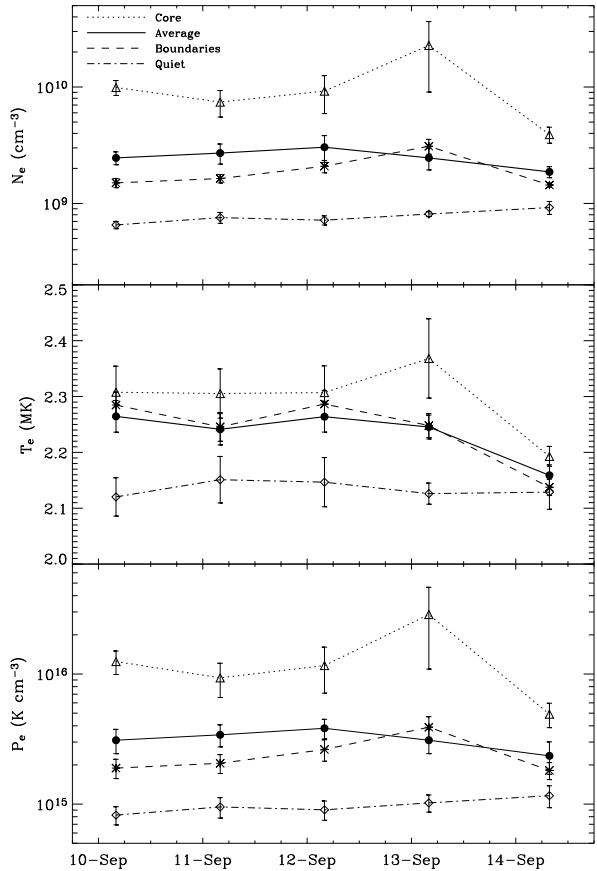}
\caption{Variations in electron density ($N_e$), temperature ($T_e$), and 
pressure ($P_e$) for September 10--14.}
\label{ne_te_pe_plot}
\end{figure}

The density maps, presented in the second row of Fig.~\ref{maps_fig}, show
values of $\sim$10$^{10}$~cm$^{-3}$  in the active region core, and
$\sim$7$\times$10$^{8}$~cm$^{-3}$ in the region boundaries. These are  in
good agreement with previous active region density measurements (Gallagher et al.~2001; Warren \&
Winebarger~2003). Densities in the core were observed to fall from $\sim$10$^{10}$
on September 10 to 3.9$\times$10$^{9}$~cm$^{-3}$ on September 14, excluding
the high density limit ($\sim$2.5$\times$10$^{10}$ cm$^{-3}$) that was reached during a C-class flare on September 13.
This behaviour is more evident in the top panel of Fig.~\ref{ne_te_pe_plot}, which clearly shows
that the average electron density changed by $\sim$30\ per cent over the five days. Subsequently, plasma that was found to reach the high density limit was excluded from any further calculations. The average electron densities are listed in
Table~\ref{res_tab}.

\begin{table*}
\footnotesize
\caption{Principal Parameters Derived for NOAA 10103 for 2002 September 10--14}
\label{res_tab}
\begin{tabular}{lccccccccccc} 
\hline
  		   & $\langle N_e \rangle$	&  $\langle T_e \rangle$   &  $\langle P_e \rangle$  	& $A$			&  $V$ 	     &  $L$ \\      
Date      	   &  10$^{9}$cm$^{-3}$ 	&  10$^{6}$K		   &  10$^{15}$K cm$^{-3}$   	& 10$^{18}$cm$^{2}$  	&  10$^{28}$cm$^{3}$  & 10$^{9}$cm \\
\hline 
2002 September 10&  2.46$\pm$0.31 		&  2.26$\pm$0.03	   &  3.10$\pm$0.60		& 12.66$\pm$0.75   	&  1.69$\pm$0.15      & 6.31$\pm$0.37 \\
2002 September 11  &  2.71$\pm$0.53  		&  2.24$\pm$0.03	   &  3.41$\pm$0.84	 	& 10.23$\pm$0.64   	&  1.23$\pm$0.11      & 5.66$\pm$0.35 \\
2002 September 12  &  3.04$\pm$0.79 		&  2.26$\pm$0.03	   &  3.83$\pm$1.16		& 10.58$\pm$0.76   	&  1.29$\pm$0.14      & 5.76$\pm$0.41 \\	   
2002 September 13  &  2.46$\pm$0.52		&  2.24$\pm$0.02	   &  3.10$\pm$0.65		&  9.89$\pm$1.03  	&  1.17$\pm$0.18      & 5.57$\pm$0.58 \\
2002 September 14  &  1.87$\pm$0.20 		&  2.16$\pm$0.02	   &  2.35$\pm$0.43		&  6.54$\pm$1.22  	&  0.62$\pm$0.17      & 4.53$\pm$0.84 \\
\hline
	            &  $P_R$		     	&  $P_C$		      	&  $P_F$  		 	&  $P_H$			 &  $\Phi_{tot}$  \\
Date	            &  10$^{25}$ergs~s$^{-1}$  	&  10$^{25}$erg~s$^{-1}$	&  10$^{25}$erg~s$^{-1}$   	&  10$^{25}$ergs~s$^{-1}$	 &  10$^{24}$Mx   \\
\hline
2002 September 10   &   3.37$\pm$0.90	     	&  0.28$\pm$0.07	      	&  0.25$\pm$0.04  	 	&  3.90$\pm$1.57		&  5.97$\pm$0.35 \\
2002 September 11   &	2.97$\pm$1.12		&  0.19$\pm$0.05		&  0.37$\pm$0.08		&  3.53$\pm$1.92		&  4.01$\pm$0.25 \\  
2002 September 12   &   3.92$\pm$2.07	     	&  0.21$\pm$0.06	      	&  0.43$\pm$0.12  	 	&  4.56$\pm$3.10		&  3.97$\pm$0.28 \\
2002 September 13   &   2.33$\pm$1.05	     	&  0.30$\pm$0.07	      	&  0.42$\pm$0.14  	 	&  2.59$\pm$1.40		&  2.78$\pm$0.29  \\
2002 September 14   &   0.71$\pm$0.24	     	&  0.07$\pm$0.03	      	&  0.18$\pm$0.06  	 	&  0.97$\pm$0.68		&  0.91$\pm$0.17 \\  
\hline
\normalsize
\end{tabular}
\end{table*}

\subsection{Electron Temperatures}
\label{temp_map}

Electron temperature maps were created using the method of
\citet{bros96}, which employs the ratio of the fluxes of various ionization stages 
of iron. This method obtains a polynomial fit to the logarithm of the temperature as a function of the logarithm of the emissivity ratio for selected line pairs, under the assumption of an isothermal plasma, with the form,
\begin{equation}
\label{brosius_eqn}
log~T_{e}=a_{0}+a_{1}(log~R)+a_{2}(log~R)^{2}+a_{3}(log~R)^{3},
\end{equation}
where the parameters, $a_{0}$, $a_{1}$, $a_{2}$, and $a_{3}$ were initially tabulated by 
\citet{bros96}, and $R$ is the intensity ratio 
\textsc{Fe xvi}~(335.41~\AA)/\textsc{Fe xiv}~(334.18~\AA).

\begin{figure}
\includegraphics[height=8.0cm, angle=90]{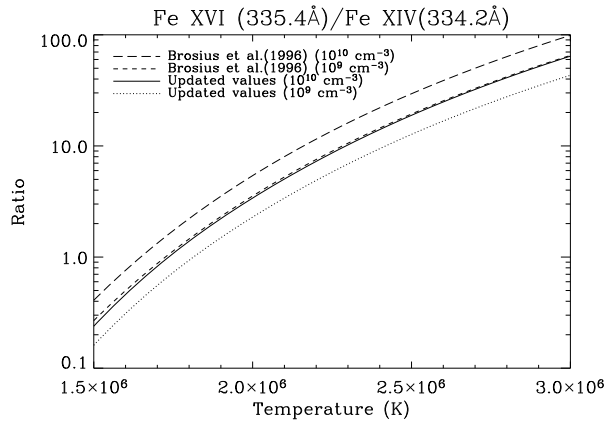}
\caption{The temperature sensitive curves for \textsc{Fe xvi} (335.41~\AA)/\textsc{Fe xiv} (334.18~\AA) for the \citet{bros96} method and the updated \textsc{chianti} v4.2 data, both plotted for plasma densities of 10$^{9}$ and 10$^{10}$ cm$^{-3}$.}
\label{bros_ch_fig}
\end{figure}

Using the most recent theoretical atomic data from \textsc{chianti} v4.2, the values of $a_{0}$,..., $a_{3}$ were found to change somewhat and resulted in a slightly higher temperature ($\sim$5\ per cent) than those predicted using \cite{bros96} values. Both previous and updated parameters have been included in Table~\ref{bros_chianti}, while the resulting temperatures are presented in Fig.~\ref{bros_ch_fig}. In addition, the \textsc{Fe xiv} (334.18~\AA) line is density sensitive, and was accounted for by determining the temperature across the active region at the corresponding density.

The temperature maps are displayed in the third row of Fig.~\ref{maps_fig}. As
expected, the temperature maps show a close spatial correlation  with the intensity
and density maps. As the region
evolves, the temperature remains constant at $\sim$2.25$\times$10$^{6}$~K, with  just a slight
decrease (by $\sim$4\ per cent) on the final day of observation. More statistically
significant is the variation of electron temperature across the region, which
ranges from $\sim$2.10$\times$10$^{6}$~K in the active region boundary to $\sim$2.34$\times$10$^{6}$~K in the core. The
values derived for the temperature are heavily dependent on the $a_{0}$ coefficient
in Equation~(\ref{brosius_eqn}), which accounts for the restricted temperature
sensitivity of this method. The temperature is also affected by the brightening on
September 13 as can be seen from the corresponding map. Average temperature
values are listed in Table~\ref{res_tab}, and are plotted in Fig.~\ref{ne_te_pe_plot}. Again, these results are
in good agreement with those found in \cite{gall01} using similar methods.

\begin{figure}
\includegraphics[height=8.0cm, angle=90]{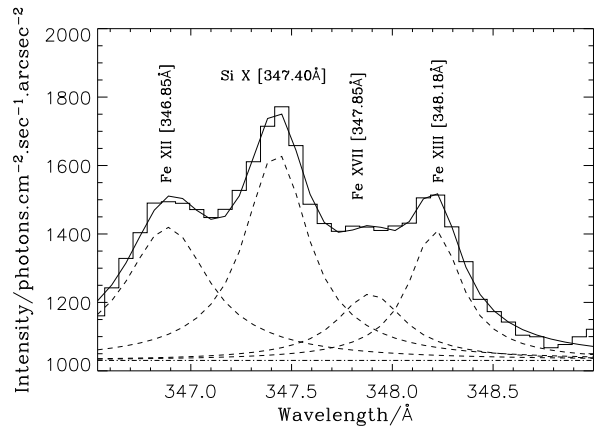}
\caption{Portion of the spectrum from the compact brightening on September 13 showing the \textsc{Fe xvii} (347.85~\AA) emission line and neighbouring lines. \textsc{Fe xvii} has a formation temperature of $\sim~5~\times~10^{6}~K$.}
\label{fe17_plot}
\end{figure}

The C-class flare on September 13 is not only evident in the intensity and density
maps of September 13, but was also identified due to the presence of the
\textsc{Fe xvii} (347.85~\AA) line which has a formation temperature of $\sim$5$\times$10$^{6}$~K. A portion of the spectrum for the flare is shown in Fig.~\ref{fe17_plot}.

\begin{table}
\footnotesize
\caption{Comparison between the parameters used in Equation~\ref{brosius_eqn} for \textsc{Fe xvi}~(335.41~\AA)/\textsc{Fe xiv}~(334.18~\AA).}
\label{bros_chianti}
\begin{tabular}{lccccc}
\hline
\multicolumn{1}{c}{Study}	&Log N$_{e}$	&$a_{0}$	&$a_{1}$	&$a_{2}$	&$a_{3}$ \\
\hline
\cite{bros96}	&9.0	&6.237		&0.111 		&0.0087  	&0.0016  \\
		&10.0	&6.217		&0.108 		&0.0076  	&0.0018  \\
Updated values	&9.0  	&6.259		&0.111 		&0.0105	 	&0.0015  \\
	 	&10.0	&6.240		&0.108 		&0.0097  	&0.0015  \\
\hline
\normalsize
\end{tabular}
\end{table}

\subsection{Electron Pressures}
\label{pres_map}

The pressure maps in the bottom row of Fig.~\ref{maps_fig} follow a similiar behaviour to that seen in the density. Values for the pressure in the high density core of the active region remain $\sim$10$^{16}$~K cm$^{-3}$ from September 10--13 and drop to 4$\times$10$^{15}$~K cm$^{-3}$ on September 14, again not taking into account the effects of the brightening. Similar to the average electron density, the average pressure varied by $\sim$30\ per cent between September 10--14 (see the values presented in Table~\ref{res_tab} and Fig.~\ref{ne_te_pe_plot}).

\subsection{Power Balance}
\label{energy_bal}
The steady-state energetics of a coronal loop can be expressed as,
\begin{equation}
\label{energy_bal_eqn}
E_H = E_R + E_C + E_F~~{\rm~ergs~cm^{-3}~s^{-1}},
\end{equation}
where $E_R$ is the total radiative losses of the plasma, $E_C$ is the thermal 
conductive flux, $E_F$ is the energy lost due to mass flows, and $E_H$ is the energy required to balance these losses (e.g. Antiochos \& Sturrock 1982, Bradshaw \& Mason 2003). 
The radiative loss term, $E_R$, in 
Equation~(\ref{energy_bal_eqn}) can be written in terms of the electron density,
$N_e$, and the radiative loss function, $\Lambda(T_e)$,
\begin{equation}
\label{rad_eqn}
E_R = -N_e^2 \Lambda(T_e)~~{\rm~ergs~cm^{-3}~s^{-1}}.
\end{equation}
The radiative loss function is usually approximated by analytical expressions of the 
form $N_e^2 \chi T_e^\alpha$ \citep{cook89}. While this method is 
simple to implement computationally, it does not capture accurately the
fine-scale structure of the radiative loss function. A more appropriate 
technique is used here, which relies on interpolating the values obtained 
from \textsc{chianti}, using the coronal abundances of \cite{feld92} and the ionization 
balance calculations of \citet{mazz98}. The choice of coronal abundances was motivated by recent work by \cite{delz03} who show that low FIP elements in active regions have abundances which are coronal.

Assuming classical heat conduction along the magnetic field, the conductive 
flux can be expressed as,
\begin{equation}
\label{conduct_eqn}
E_C = \frac{d}{ds} \left [ -\kappa T_e^{5/2} 
\frac{dT_e}{ds} \right ]~~{\rm~ergs~cm^{-3}~s^{-1}},
\end{equation}
where the thermal conductivity is $\kappa = 0.92 \times 10^{-6}~~{\rm 
ergs~s^{-1}~cm^{-1}~K^{-7/2}}$ \citep{spit62}, $T_e$ is the electron temperature. Various approximations for Equation~(\ref{conduct_eqn}) have been used in previous studies: \citet{asch99} used an expression that was heavily dependent upon the temperature gradient, $(dT/ds)^2$, in the study of active region loops, whereas \cite{vara00} used an approximation strongly dependent upon temperature, $T^{7/2}$, for post-flare loops. Here, we assume a semi-circular loop geometry and approximate Equation~(\ref{conduct_eqn}) by,
\begin{equation}
\label{conduct_approx}
E_C = -\kappa T_e^{5/2} \frac{\Delta T_e}{(L/2)^{2}}~~{\rm~ergs~cm^{-3}~s^{-1}}, 
\end{equation}
where $\Delta T_{e}$ is the difference in the maximum and minimum temperatures given in Section~\ref{temp_map}, and $ds$ is approximated by $L/2$. This approximation is therefore not overly dependent on temperature or the temperature gradient, but acts as a balance between the approximations of \cite{asch99} and \cite{vara00}. 

The energy losses due to mass motions of the plasma, $E_F$, can be expressed as a sum of the kinetic energy and the internal energy of the plasma,
\begin{equation}
\label{flow_eqn}
E_F = \frac{d}{ds} \left [ v \left ( \frac{1}{2} \rho v^2 + \frac{5}{2} N_e k_B T_e \right ) \right ]~{\rm~ergs~cm^{-3}~s^{-1}},
\end{equation}
where $v$ is the flow velocity of the plasma, and $ds$ is again approximated to be the loop half-length ($L/2$). Line-of-sight velocities at the core of the active region were found to be $\sim$10~km~s$^{-1}$ which is consistant with \citet{bryn99} who detected flow speeds of $\sim$15~km~s$^{-1}$ also using CDS. Using these velocity values, as well as typical parameters from Table~\ref{res_tab}, the mass loss term was found to be $\sim$10$^{24}$~ergs~s$^{-1}$. Energy losses due to mass motions can therefore be considered negligible compared to radiative losses.

\begin{figure}
\includegraphics[width=8.0cm]{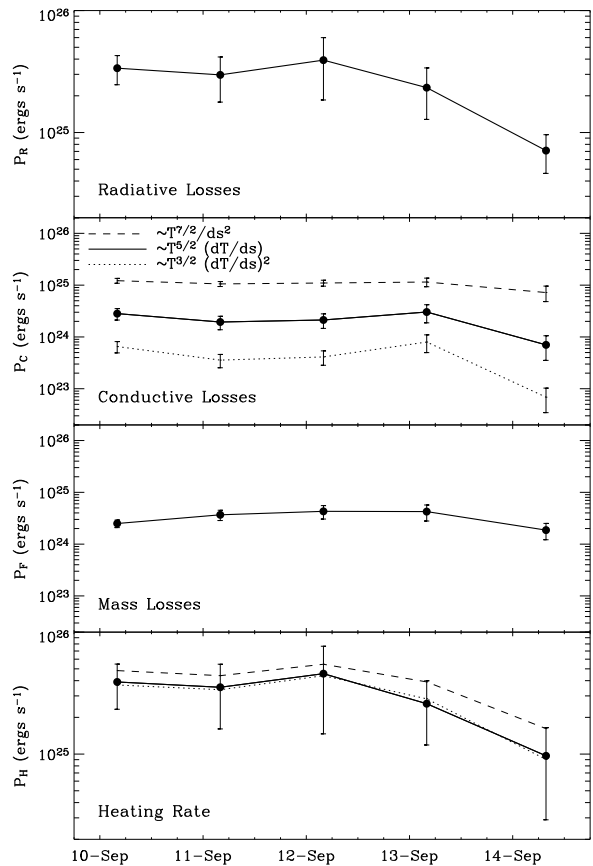}
\caption{The radiative, conductive, and mass losses, and the heating rate 
for NOAA 10103 for September 10--14.}
\label{plot_loss}
\end{figure}

In order to compare the work described in this paper with that of 
\citet{fish98}, \citet{park83}, and others, Equation~(\ref{energy_bal_eqn}) 
has been integrated over the volume of the active region, and expressed as 
a power balance equation:
\begin{equation}
\label{power_eqn}
P_H = P_R + P_C + P_F ~~{\rm~ergs~s^{-1},}
\end{equation}
where $P_R$, $P_C$, and $P_F$ are the power lost due to radiation, conduction, and mass flows, 
respectively. $P_H$ is then the heating rate required to balance these losses.

Using the active region properties detailed in Table~\ref{res_tab}, the
radiative, conductive, and mass flow losses were then calculated using
Equations~(\ref{rad_eqn}), (\ref{conduct_approx}), and (\ref{flow_eqn}). These
results, together with the heating rate calculated using
Equation~(\ref{power_eqn}), are presented in Table~\ref{res_tab} and
Fig.~\ref{plot_loss}. Due to the difficulty in correlating emission seen in CDS
with a particular magnetic flux concentration, the region's average properties
were used for comparision with MDI. The second panel down of
Fig.~\ref{plot_loss} also shows how the conductive flux values depend on how
Equation~(\ref{conduct_eqn}) is approximated. In each case, the conductive
losses are found to be much less than the radiative losses. Indeed, the average
radiative losses ($\langle P_R \rangle \sim 3 \times 10^{25}$~ergs~s$^{-1}$)
are found to exceed both the conductive losses ($\langle P_C \rangle \sim 2
\times 10^{24}$~ergs~s$^{-1}$) and the mass flow losses ($\langle P_F \rangle
\sim 4 \times 10^{24}$~ergs~s$^{-1}$) by approximately an order of magnitude.
\cite{asch00} also found an order of magnitude difference between the radiative
and conductive losses in coronal loops despite the actual values being
somewhat  lower than those found here due to the insensitivity of EIT filter
ratio techniques. As can be seen from Fig.~\ref{plot_loss}, the heating rate
falls by close to a factor of 5 between September 10 and 14 and is not
significantly affected by whichever conductive loss equation is used. 

\begin{figure}
\includegraphics[height=8.0cm, angle=90]{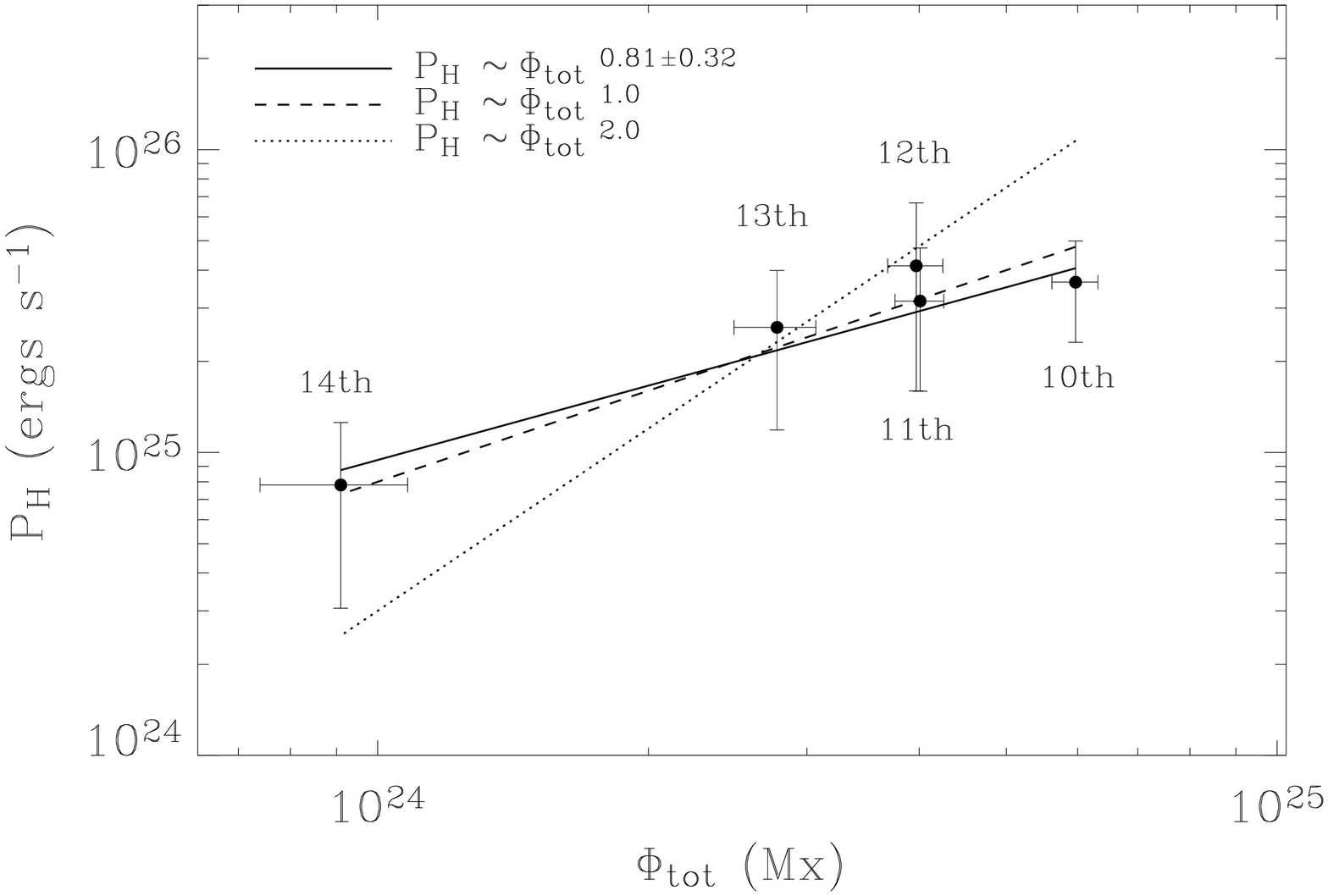}
\includegraphics[height=8.0cm, angle=90]{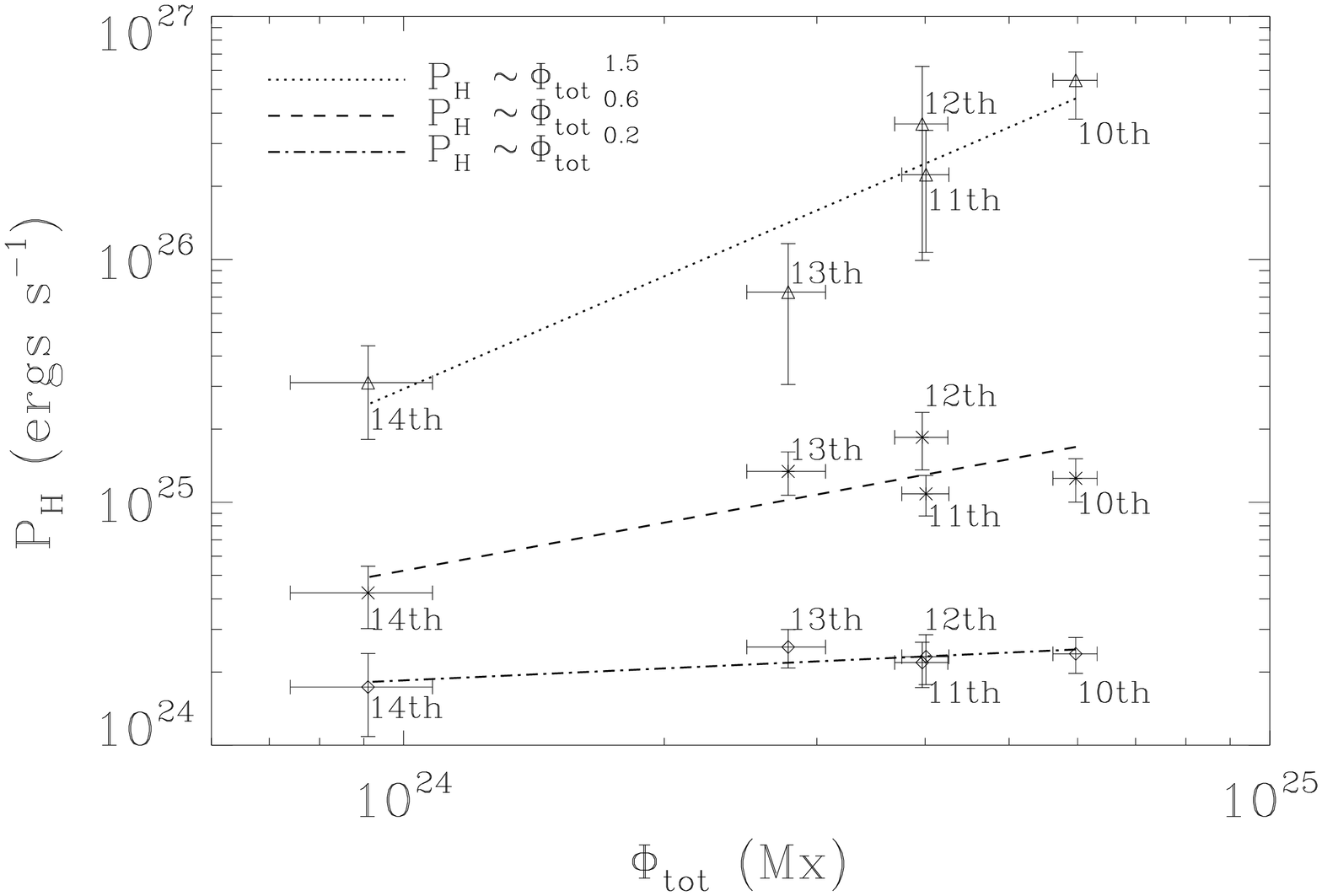}
\caption{{\it Top Panel}: Plot of $P_H$ versus $\Phi_{tot}$ for five days of MDI
and CDS observations. Also shown is a least-squares fit to the non-flaring data
together with power-laws with slopes of 1 and 2 from the theoretical
predictions. {\it Bottom Panel}: Plot of $P_H$ versus $\Phi_{tot}$, where $P_H$
has been calculated using non-flaring core ({\it dotted line}), boundary ({\it
dashed line}), and quiet-Sun ({\it dot-dashed line}) parameters from
Fig.~\ref{ne_te_pe_plot}.}
\label{phi_power}
\end{figure}

The top panel of Fig.~\ref{phi_power} shows the heating rate as a function of
the total unsigned magnetic flux. A least-squares fit to the non-flaring data
(i.e., neglecting the high density plasma of September 13) yielded a power-law
of the form $P_H \sim \Phi_{tot}^{\gamma}$, where $\gamma = 0.81\pm0.32$.
Power-laws with slopes of 1 and 2 from Equations~(\ref{ac_heating}) and
(\ref{dc_heating}) are also shown for comparison. The bottom
panel of Fig.~\ref{phi_power} shows how this relationship varies when densities from the core and the boundaries of the active region as well as from the quite solar corona (see Fig.~\ref{ne_te_pe_plot}) are used instead of average values. This wide range of $N_{e}$ shows that ${\gamma}$ varies from 0.2 in the quiet-Sun to 1.5 at the core of the region. It can therefore be concluded that average values of these parameters are a reasonable representation of the entire active region.

\section{DISCUSSION AND CONCLUSIONS}
\label{conc}

A detailed study of an evolving active region has been described using
measurements from several instruments onboard {\it SOHO}. The region was
observed to decay in size and complexity as it passed from close to the central
meridian on September 10, to close to the west limb on September 14. In the
photosphere, the total sunspot area fell by close to a factor of 2, while the
total magnetic flux fell by approximately a factor of 6. In the corona, the
average electron density, temperature, and pressure all showed similar decreases
in value, which is to be expected considering that the plasma in the corona
traces out field lines which are ultimately rooted in the photosphere. These
results are of consequence to efforts in understanding active region evolution
and the relationship between the photosphere and corona of solar active regions
(e.g., Abbett \& Fisher 2003; Ryutova \& Shine 2004).

In addition to studying active region evolution, CDS and MDI were used to 
investigate the power-law relationships predicted by theoretical models of the
corona. \citet{mand00} also investigated theoretical scaling laws using magnetic
field extrapolations from both vector and longitudinal magnetograms, in light
of the work of \citet{klim95}. They concluded that models involving the gradual
stressing of the coronal magnetic field are in better agreement with the
observational contrains than are wave heating models. This same general
conclusion was also reached by \cite{demoulin03} using photospheric and coronal
measurements from MDI and {\it Yohkoh}. Unfortunately these studies relied on
broad-band filter ratios to determine electron temperatures and densities; the
difficulty associated with making such measurements is clear from the
contradictory results of \citet{priest98}, \citet{asch01}, and \citet{reale02},
who also investigated coronal loop heating models using filter ratios from {\it
Yohkoh}/SXT. 

The analysis presented in this paper, on the other hand, is based
on well understood line ratio techniques, which offer a less ambiguous
determination of plasma properties. With this in mind, the power-law
relationship between the total heating rate and the total unsigned magnetic flux
($\Phi_{tot}$) was determined, finding a relationship of the form $P_H \sim
\Phi_{tot}^{0.81\pm0.32}$. \cite{fish98} compared the X-ray
luminosity (which was assumed to be some fraction of the total heating power) to
active region vector magnetograms, finding a similar power-law relation of $L_{X}
\sim \Phi_{tot}^{1.19}$. The result of \cite{fish98} suggests a ``universal''
relationship between magnetic flux and the amount of coronal heating, regardless
of the age or complexity of the active region. A similar relationship of $L_X \sim
\Phi_{tot}^{0.9}$ was found using statistical samples of late-type stars
\citep{schr87}. The results of this paper therefore lend further observational evidence
that active regions are heated by magnetically-associated waves, rather
than multiple nanoflare-type events.

\section*{Acknowledgments}
{\it SOHO} is a project of international collaboration between ESA and NASA. 
This work has been supported by a Cooperative Award in Science and Technology
(CAST) studentship from Queen's University Belfast and the NASA GSFC {\it SOHO} project. ROM would like to thank R. T. J. McAteer for useful comments. FPK is grateful to AWE Aldermaston for the award of a William Penny Fellowship.

\end{document}